\documentclass[usenatbib,useAMS]{mn2e}
\usepackage[fleqn]{amsmath}
\usepackage{txfonts}
\usepackage{graphicx}
\usepackage{hyperref}
\DeclareSymbolFont{cmletters}{OML}{cmm}{m}{it}
\DeclareMathSymbol{v}{\mathalpha}{cmletters}{"76}
\defcitealias{Sari1995}{SP95}
\defcitealias{Kennel1984}{KC84}
\defcitealias{Panaitescu2002}{PK02}
\newcommand{\os}[1]{o\!\left(#1\right)}
\title[Constraints on Cold Magnetized Shocks in GRBs]{Constraints on Cold Magnetized Shocks in Gamma-Ray Bursts}

\author[R. Narayan, P. Kumar, and A. Tchekhovskoy]{Ramesh
  Narayan$^{1}$\thanks{E-mail: rnarayan@cfa.harvard.edu (RN);
    pk@astro.as.utexas.edu (PK); atchekho@princeton.edu (AT)}, 
  Pawan Kumar$^{2}$\footnotemark[1] and Alexander
  Tchekhovskoy$^{3}$\footnotemark[1]\thanks{Princeton Center for
    Theoretical Science Fellow}\\
$^{1}$Harvard-Smithsonian Center for Astrophysics, Harvard University, 60 Garden Street, Cambridge, MA 02138, USA\\
$^{2}$Astronomy Department, University of Texas, Austin, TX 78712, USA\\
$^{3}$Princeton Center for Theoretical Science, Jadwin Hall, Princeton University, Princeton, NJ 08544, USA}

\begin{document}

\date{Accepted 2011 June 6. Received 2011 April 29}

\pagerange{\pageref{firstpage}--\pageref{lastpage}} \pubyear{2011}

\maketitle

\label{firstpage}

\begin{abstract}
We consider a model in which the ultra-relativistic jet in a gamma-ray
burst (GRB) is cold and magnetically accelerated. We assume that the
energy flux in the outflowing material is partially thermalized via
internal shocks or a reverse shock, and we estimate the maximum amount
of radiation that could be produced in such magnetized shocks. We
compare this estimate with the available observational data on prompt
$\gamma$-ray emission in GRBs. We find that, even with highly
optimistic assumptions, the magnetized jet model is radiatively too
inefficient to be consistent with observations. One way out is to
assume that much of the magnetic energy in the post-shock, or even
pre-shock, jet material is converted to particle thermal energy by
some unspecified process, and then radiated.  This can increase the
radiative efficiency sufficiently to fit observations. Alternatively,
jet acceleration may be driven by thermal pressure rather than
magnetic fields.  In this case, which corresponds to the traditional
fireball model, sufficient prompt GRB emission could be produced
either from shocks at a large radius or from the jet photosphere
closer to the center.
\end{abstract}


\section{Introduction}\label{intro}

A great deal of progress has been made in our understanding of
gamma-ray bursts (GRBs), thanks to the launch of a number of dedicated
satellites (BeppoSAX, HETE-2, Swift and Integral).  These satellites
rapidly communicate burst locations to ground-based optical and radio
telescopes, which has enabled detailed follow up study of the GRB
afterglow emission. It is now known that GRBs produce highly
relativistic and beamed jets containing energy $\sim10^{51}$\,erg \citep*[see][for extensive reviews of these and other developments]{Meszaros2002, Piran2005, Zhang2007,Gehrels2009}.  It is
also well established that there are two classes of GRBs.  One class,
called long-GRBs --- those lasting for more than a few seconds --- is
produced when a massive star collapses at the end of its nuclear
burning life (see \citealt{Woosley2006} for a review). For the other
class, called short-GRBs -- those lasting for less than a few seconds
-- at least some members are believed to result from mergers of
compact stars in binary systems (\citealt{Gehrels2009}, and references
therein).

Despite this impressive progress, several fundamental questions remain
unanswered.  Foremost among these is the composition of the
relativistic jets that power GRBs. We do not know whether GRB jets
consist of a normal proton-electron plasma or if they are dominated
by electron-positron pairs. Furthermore, it is uncertain whether the
jets are dominated by matter or magnetic fields (Poynting outflow).
The related question of how the observed $\gamma$-ray radiation is
produced is also poorly understood.

A popular model for converting jet energy to particle thermal energy
and thereby to radiation is the internal shock model \citep{Narayan1992,Rees1994,Sari1997}.  According to
this model, the relativistic wind from the central engine of a GRB has
a variable Lorentz factor, which leads to collisions between faster
and slower moving ejecta. A fraction of the kinetic energy of the jet
is converted to thermal energy in these ``internal'' shocks. A
fraction of this thermal energy then goes into electrons and is
rapidly radiated away as $\gamma$-ray photons via synchrotron and
inverse-Compton processes. The internal shock model naturally produces
the rapid variability observed in the $\gamma$-ray emission of GRBs
\citep{Sari1997}. This is one of its principal virtues.

The internal shock model, however, has a problem, viz., the efficiency
$\epsilon_\gamma$ (see eq. \ref{egamma} for the definition) for
converting jet energy to radiation is relatively low \citep{Kumar1999,
Lazzati1999,Panaitescu1999}. The efficiency depends on the relative Lorentz factor of the
colliding blobs, and also, in the case of magnetized ejecta, on the
jet magnetization parameter $\sigma$ (defined in eq. \ref{sigma}).
Since the efficiency $\epsilon_\gamma$ of a GRB can be measured
directly from observations of the prompt and afterglow emission, we
can constrain the parameters of the internal shock model, notably the
magnetization $\sigma$ of the jet material.

Another location where the jet energy may possibly be converted to
$\gamma$-rays might be the deceleration radius where the jet starts to
slow down as a result of its interaction with an external medium.  Two
shocks are formed in this interaction, one of which, the ``forward''
shock, heats up the external medium and produces the afterglow
emission, and the other, the ``reverse'' shock, propagates into the
GRB jet. The energy released in the reverse shock could be radiated as
$\gamma$-rays via synchrotron and/or inverse-Compton
processes.{\footnote{It would be very difficult to produce the
observed $\gamma$-ray variability in the reverse shock model unless
there is relativistic turbulence in the shocked fluid \citep{NarayanKumar2009,Lazar2009}.}  The efficiency for
converting jet energy to $\gamma$-rays depends on various parameters,
including the $\sigma$ of the jet material.

Thus, in either the internal shock model or the reverse shock model,
the $\gamma$-ray efficiency $\epsilon_\gamma$ depends on the
magnetization $\sigma$ of the jet.  The present work is motivated by
the fact that, under some circumstances, we can independently estimate
$\sigma$ for a GRB jet.  This follows from the recent work of
\citet*{Tchekhovskoy2010b} who studied the properties of
a magnetically accelerated GRB jet.  If the jet material is cold,
i.e., there is no thermal pressure, and all the acceleration is from
electromagnetic forces (Poynting-dominated jet), these authors show
that $\sigma$ can be estimated from the terminal Lorentz factor
$\gamma_j$ and the opening angle $\theta_j$ of the jet. Both of the
latter quantities can be measured from afterglow data.  We thus have
an opportunity to check if the values of $\sigma$ obtained from
observations of GRB afterglows are consistent with the $\gamma$-ray
efficiencies $\epsilon_\gamma$ measured for the same GRBs. Carrying
out this test is our goal.

In \S\ref{sec:relativisticperpendicularshock} we write down the standard jump conditions for a magnetized
relativistic ``perpendicular'' shock in which the magnetic field is
perpendicular to the flow velocity (or parallel to the shock
front). By solving the jump conditions, we calculate the efficiency
with which the kinetic energy of a cold relativistic magnetized fluid
is converted by the shock to thermal energy.  In \S\ref{sec:internalshockmodel}, we calculate the
efficiency of the internal shock model and compare it against
observations, and in \S\ref{sec:reverseshockmodel}, we carry out a similar exercise for the
reverse shock model. In both cases, we show that there is an
inconsistency between the predictions of the model and measured values
of $\epsilon_\gamma$ and $\sigma$.  We discuss the implications of
this result in \S\ref{sec:summaryanddiscussion} and suggest possible solutions.

\section{Relativistic Perpendicular Shock}
\label{sec:relativisticperpendicularshock}
\subsection{Preliminaries}

The problem of interest was discussed in detail by \citet[hereafter
\citetalias{Kennel1984}]{Kennel1984}. We follow their methods with a few minor
changes.  We consider a cold magnetized fluid with a magnetic field
strength $B_0$ in its rest frame. We assume ideal magnetohydrodynamics
(MHD) and set the
electric field in the rest frame to zero.  Transforming to a frame in
which the magnetized fluid moves with dimensionless velocity
$\beta=v/c$, Lorentz factor $\gamma=1/\sqrt{1-\beta^2}$, in a
direction perpendicular to the magnetic field, the magnetic and
electric fields become
\begin{equation}
B=\gamma B_0, \qquad E=uB_0=\frac{u}{\gamma}B,
\qquad u=\beta\gamma=(\gamma^2-1)^{1/2},
\end{equation}
where $u$ is the relativistic $4$-velocity. The fields $B$ and $E$ in the
new frame are parallel and perpendicular, respectively, to $B_0$, and
each is also perpendicular to the velocity.

We define the magnetization parameter $\sigma$ of the moving fluid as
the ratio of the Poynting energy flux to the particle rest energy
flux. Thus
\begin{equation}
\sigma=\frac{cEB/4\pi}{n\gamma umc^3}=\frac{B^2/4\pi}{n\gamma^2mc^2}
=\frac{B_0^2/4\pi}{nmc^2}, \label{sigma}
\end{equation}
where $n$ is the particle number density in the fluid rest frame and
$m$ is the mass of each particle.  In the final expression, the
numerator is the rest frame ``enthalpy'' of the magnetic field, which
is equal to $[\Gamma_B /(\Gamma_B-1)](B_0^2/8\pi)$ (taking the
adiabatic index of the magnetic field $\Gamma_B=2$ for compression
transverse to the field), and the denominator is the rest energy
density. Since we see that $\sigma$ depends only on rest frame
quantities, it is a relativistic invariant and is frame-independent.
\citetalias{Kennel1984} use a slightly different definition of $\sigma$ where they
replace $\gamma^2$ in the denominator of the third quantity in
equation (\ref{sigma}) by $\gamma u$. As a result, their $\sigma$ is
not truly frame-independent.  However, the difference between the two
definitions is negligibly small for highly relativistic flows.

\subsection{Jump Conditions}

\label{sec:jumpconditions}
We follow \citetalias{Kennel1984}, except that we use the definition of $\sigma$ given in
equation (\ref{sigma}) and avoid certain approximations.  We use
subscript $u$ for the gas upstream of the shock and subscript $d$ for
the downstream gas. The upstream gas is cold ($P_u=0$), has a
magnetization parameter $\sigma$, rest number density $n_u$, and moves
with Lorentz factor $\gamma_u$ in the frame of the shock.  The
downstream gas is hot ($P_d \neq 0$) with adiabatic index $\Gamma$,
has number density $n_d$, and Lorentz factor $\gamma_d$.  In the shock
frame, the magnetic fields in the two regions, $B_u$ and $B_d$, are
related by
\begin{equation}
B_d=\frac{\gamma_d}{u_d}E_d=\frac{\gamma_d}{u_d}E_u
=\frac{\gamma_d u_u}{\gamma_u u_d}B_u,
\end{equation}
where we have used the fact that the electric field is continuous
across the shock.

The upstream and downstream gas enthalpy per particle are,
respectively,
\begin{equation}
\mu_u=mc^2, \qquad \mu_d=mc^2\left[1+\frac{\Gamma}{(\Gamma-1)}\,
\frac{P_d}{n_d mc^2}\right].
\end{equation}
The second term inside the square brackets is a dimensionless number
which describes the thermal enthalpy per particle of the shocked
gas. It can be written as
\begin{equation}
\frac{\Gamma}{(\Gamma-1)}\,\frac{P_d}{n_d mc^2} = h(\theta_d)\theta_d,
\quad \theta_d = \frac{P_d}{n_d mc^2} = \frac{kT_d}{mc^2},
\quad h(\theta_d) = \frac{\Gamma}{\Gamma-1},
\end{equation}
where $\theta_d$ is the relativistic temperature of the downstream
gas. When $\theta_d\ll1$, the gas is non-relativistic, and we have
$\Gamma=5/3$, $h(\theta_d)=5/2$, whereas when $\theta_d\gg1$, the gas
is ultra-relativistic, and we have $\Gamma=4/3$, $h(\theta_d)=4$. At
intermediate temperatures ($\theta_d \sim 1$), $h(\theta_d)$ can be
written in terms of modified Bessel functions (see \citealt{Chandrasekhar1960}).  For simplicity, we use the following approximation \citep{Service1986},
\begin{equation}
h(\theta) = \frac{10+20\,\theta}{4+5\,\theta},\label{htheta}
\end{equation}
which is sufficiently accurate for our purposes.  In principle, if the
jet material consists of a normal electron-proton plasma, we should
allow for two species of particles in the shocked gas, each with a
different temperature.  We ignore this complication for simplicity.

We have three jump conditions across the shock, corresponding to three
fundamental conservation laws. First, mass conservation implies that
the mass fluxes on the two sides of the shock must be equal, i.e.,
\begin{equation}
n_u u_u=n_d u_d.\label{mass}
\end{equation}
Energy conservation requires the energy fluxes to be equal, i.e.,
\begin{equation}
n_u u_u\gamma_u\mu_u+\frac{EB_u}{4\pi}=n_d u_d\gamma_d\mu_d+\frac{EB_d}{4\pi}.
\label{energy}
\end{equation}
Finally, momentum conservation gives the condition
\begin{equation}
n_u u_u^2\mu_u+\frac{(B_u^2+E_u^2)}{8\pi}=n_d u_d^2\mu_d+P_d
+\frac{(B_d^2+E_d^2)}{8\pi}.\label{mmtm}
\end{equation}
In the last equation, the terms involving the electric field cancel
since $E_u=E_d$.  Eliminating $\mu_d$ between equations (\ref{energy})
and (\ref{mmtm}) and simplifying, we obtain the following expression
for $\theta_d$:
\begin{equation}
\theta_d=\left[u_d^2\left(\frac{u_u}{u_d}-\frac{\gamma_u}{\gamma_d}\right)
(1+\sigma)+\left(\frac{\gamma_u^2u_d}{2u_u}-\frac{\gamma_d^2u_u}{2u_d}
\right)\sigma\right].\label{thetad1}
\end{equation}
In addition, equation (\ref{energy}) can be rewritten in the following
simplified form,
\begin{equation}
1+h(\theta_d)\theta_d = \frac{\gamma_u}{\gamma_d}
(1+\sigma)-\frac{u_u}{u_d}\sigma.\label{thetad2}
\end{equation}

Given the upstream quantities $u_u$ and $\sigma$, it is
straightforward to solve equations (\ref{thetad1}) and (\ref{thetad2})
numerically.  We guess a value for the downstream velocity $u_d$ and
calculate $\theta_d$ using equation (\ref{thetad1}).  We then compute
$h(\theta_d)$ using the approximation (\ref{htheta}) and check whether
the condition (\ref{thetad2}) is satisfied. If it is not, we
numerically adjust $u_d$ until the condition is satisfied. We then
have the complete solution for all downstream quantities: $\gamma_d$,
$u_d$, $n_d/n_u$, $\theta_d$, $\mu_d/mc^2$, $B_d/B_u$.

The results presented in the following sections use the above
numerical approach to solve the jump conditions. An alternate approach
is to make suitable approximations and obtain analytical solutions of
the jump conditions.  Appendix \ref{appendixa} presents analytical solutions
corresponding to a number of useful limits.

\section{Internal Shock Model}
\label{sec:internalshockmodel}
\subsection{Solving the Jump Conditions}

We consider two identical blobs, each with magnetization $\sigma$,
approaching each other and colliding. In the center of mass frame, the
blobs have Lorentz factors $\gamma$ and relativistic velocities
$u=\pm\sqrt{\gamma^2-1}$.  As a result of the collision, two identical
shocks move (in opposite directions) into the two blobs.

For given values of $\gamma$ and $\sigma$, we solve the jump
conditions numerically and calculate all quantities of interest in the
shocked gas. We begin by assuming a value for the upstream Lorentz
factor $\gamma_u$ in the frame of one of the shocks. Following the
procedure described in \S\ref{sec:jumpconditions}, we solve for the downstream Lorentz
factor $\gamma_d$. From $\gamma_u$ and $\gamma_d$, we calculate the
relative Lorentz factor $\gamma_{ud}$ between the two regions
(relativistic velocity subtraction) and check whether it corresponds
to the desired value of $\gamma$. If not, we adjust $\gamma_u$ until
we obtain $\gamma_{ud}=\gamma$. We then have the solution.

Having obtained the solution, we switch to the rest frame of the
shocked gas.  We assume that a fraction $\epsilon_e$ of the thermal
enthalpy of the shocked gas $W_{\rm gas}$ goes into electrons and that
it is entirely radiated in $\gamma$-rays\footnote{This is perhaps a
  little optimistic. It is possible that only the gas internal energy
  is radiated, which is $1/\Gamma$ times the enthalpy.}. This gives
the energy $E_\gamma$ that goes into $\gamma$-rays.  The remaining
unradiated enthalpy, which consists of rest mass enthalpy $W_{\rm
  rest}$, remaining gas thermal enthalpy $(1-\epsilon_e)W_{\rm gas}$
and magnetic enthalpy $W_B$, contributes to the kinetic energy $E_0$
that goes into the afterglow.  Thus, we estimate the efficiency
$\epsilon_\gamma$ of $\gamma$-ray emission, the fraction of the total
energy that goes into $\gamma$-rays, to be
\begin{equation}
\epsilon_\gamma \equiv \frac{E_\gamma}{E_\gamma+E_0} =
\frac{\epsilon_e W_{\rm gas}}{W_{\rm gas}+W_{\rm rest}+W_B}.\label{egamma}
\end{equation}

The quantities $W_{\rm gas}$ and $W_{\rm rest}$ are easily obtained
from the shock solution. Per particle, they are given by
\begin{equation}
W_{\rm gas}=h(\theta_d)\theta_d mc^2, \qquad W_{\rm rest}=mc^2.
\end{equation}
To calculate $W_B$, we first need to calculate the magnetic field of
the shocked gas in the center of mass (CM) frame of the colliding
blobs. This is given by
\begin{equation}
B_{\rm CM} = \frac{B_d}{\gamma_d} = \frac{u_u}{\gamma_uu_d}B_u.
\end{equation}
Then the magnetic enthalpy per particle is
\begin{equation}
W_B = \frac{B_{\rm CM}^2}{4\pi n_d} = \frac{u_u}{u_d}\sigma mc^2.
\end{equation}
Thus we obtain
\begin{equation}
\epsilon_\gamma =
\frac{\epsilon_e h(\theta_d)\theta_d}{1+h(\theta_d)\theta_d+(u_u/u_d)\sigma}.
\end{equation}
Although all quantities have been estimated in the rest frame of the
shocked gas, it is easily shown that Lorentz transforming to a
different frame, e.g., the observer frame, will leave
$\epsilon_\gamma$ unchanged.

\begin{figure}
\includegraphics[width=\columnwidth]{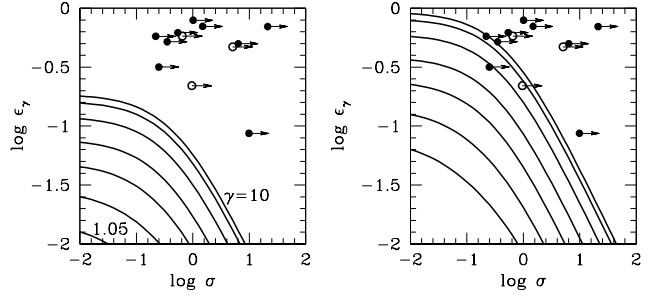}
\caption{Left: $\gamma$-ray efficiency $\epsilon_\gamma$ vs the
upstream magnetization parameter $\sigma$ for an internal shock
between two identical blobs. It is assumed that a fraction
$\epsilon_e=0.2$ of the enthalpy in the shocked gas is radiated as
prompt $\gamma$-rays.  From below, the curves correspond to blob
Lorentz factor in the center-of-mass frame of $\gamma=1.05$, 1.1, 1.2,
1.4, 2, 4, 10, respectively. Equivalently, the inter-blob Lorentz
factor between the two blobs is $\gamma_{\rm ib}=1.21$, 1.42, 1.88,
2.92, 7, 31, 199, respectively.  The symbols refer to observational
data. The filled circles correspond to the first nine GRBs in Table \ref{tab1}
(\citetalias{Panaitescu2002} sample) and the open circles to the last three GRBs.  The arrows
on the symbols indicate that the estimated values of $\sigma$ are
lower limits (eq. \ref{sigmamin}).  Right: Corresponding results for the case when
$\epsilon_e=1$, i.e., all the thermal enthalpy in the shocked gas
comes out in prompt $\gamma$-rays.}
\label{fig1}
\end{figure}

For a given value of $\epsilon_e$, the $\gamma$-ray efficiency depends
on two parameters, the magnetization $\sigma$ of the blobs and their
Lorentz factor $\gamma$ in the center-of-mass frame.  Figure \ref{fig1} shows
how $\epsilon_\gamma$ varies as a function of $\sigma$ for selected
values of $\gamma$, as listed in the figure caption.  Instead of
$\gamma$, we could express the results in terms of the relative
inter-blob Lorentz factor of the blobs $\gamma_{\rm ib}$. The values
of $\gamma_{\rm ib}$ are also given in the figure caption.

Figure \ref{fig1} indicates that the maximum radiative efficiency is obtained
for unmagnetized blobs.  As the magnetization increases, the amount of
thermal energy generated in the shock decreases, reducing the
radiative efficiency. This result was already discussed by \citetalias{Kennel1984}.  The
analytical approximations in Appendix \ref{appendixa} give further details. From
\S\ref{sec:A3} we see that, for $\sigma\ll1$, the enthalpy of the shocked gas
(which is proportional to $\theta_d$) varies as $1-(5/2)\sigma$, i.e.,
it reduces with increasing magnetization. The reduction is quite
pronounced once $\sigma>1$; for $\sigma\gg1$, the enthalpy scales as
$1/\sigma$ (\S\ref{sec:A1}).

Note, however, that there is always a shock solution for any choice of
$\gamma$ and $\sigma$.  This may appear a little surprising since, as
we show in Appendix \ref{appendixa}, a magnetized shock is possible only if the
upstream velocity $u_u$ exceeds $\sqrt{\sigma}$.  Thus, for strongly
magnetized blobs moving in the center-of-mass frame with relatively
low velocities, one might think there should be no shock. For example,
for $\gamma=10$ or $u=9.95$, one might expect the shock to disappear
once $\sigma>3.15$, whereas Fig. \ref{fig1} shows results for $\sigma$ values
much above this limit.

The explanation is simple. What is relevant for the existence or
otherwise of a shock is the upstream velocity in the {\it shock
frame}, not the center-of-mass frame.  In all our solutions, the shock
moves outward in the center-of-mass frame. When $u^2\gg\sigma$, the
shock velocity is not very large. However, once $\sigma$ exceeds
$u^2$, the shock moves quite rapidly into the blob. In fact, it moves
so rapidly that the velocity of the upstream gas $u_u$ as seen in the
shock frame becomes larger than $\sqrt{\sigma}$, thus permitting a
shock.  Shocks in this regime are however weak (see \S\ref{sec:A2}), and
their radiative efficiencies $\epsilon_\gamma$ are correspondingly
very low. As an example, note in Fig. \ref{fig1} the very low efficiency of the
$\gamma=10$ model (the uppermost curves in the two panels) when
$\sigma=10$.

\subsection{Comparison with GRB Data}\label{data}

\begin{table*}
\begin{center}
\caption{GRB data (see \S\ref{data} for details). }
\begin{minipage}{1.2\columnwidth}
\begin{tabular}{lccccccc}
\hline
Source &  ${E_\gamma}^{a}$ & ${E_0}^{b}$ & ${\epsilon_\gamma}^{c}$ & $\gamma_{j}$ & $\theta_j(^\circ)$ & $\sigma_{\rm min}$ &  Ref. \\
\hline
GRB 970508 &  3.8 &    20.   &    0.087   &     150    &     18.3   &        9.86  &   1,2 \\
GRB 990123 &  4.9 &   1.5    &   0.62     &    300     &     2.1    &       0.54   &   1   \\
GRB 990510 &  1.3 &   1.4    &   0.32     &    140     &     3.1    &       0.25   &   1   \\
GRB 991208 &   18 &   2.4    &   0.79     &     68     &    12.8    &       1.01   &   1   \\
GRB 991216 &  3.0 &   1.1    &   0.58     &    150     &     2.7    &       0.22   &   1   \\
GRB 000301c&  6.6 &   3.3    &   0.50     &    160     &    13.7    &       6.38   &   1   \\
GRB 000418 &  148 &   32     &   0.70     &     90     &    50.0    &      21.1    &   1   \\
GRB 000926 &   15 &   3.2    &   0.70     &    130     &     8.1    &       1.49   &   1   \\
GRB 010222 &   11 &   5.1    &   0.52     &    110     &     4.6    &       0.35   &   1   \vspace{5pt}\\
GRB 021004 &  560 &   400    &   0.58     &     55     &    12.7    &       0.65   &  2,3   \\
GRB 080916C& $8.8\times10^4$&$10^5$& 0.47 &   $>880$   &    $>2.2$  &     $>5.07$  & 4,5,6  \\
GRB 090510 & 1100 &  4000    &   0.22     &  $>1200$    &     0.7    &     $>0.96$   &  7,8   \\
\hline    
\label{tab1}
\vspace{-8pt}
\end{tabular}\\
$^a$ Beaming-corrected energy in $\gamma$-rays in units of $10^{50}$\,erg\\
$^b$ Beaming-corrected kinetic energy in the afterglow in units of $10^{50}$\,erg\\
$^c$ Calculated using eq. (\ref{egamma}) for the last three GRBs, but with $E_0$ replaced by $2E_0$ for the first nine GRBs\\
References: 1 -- \citetalias{Panaitescu2002}; 2 -- \citet{Bloom2003};
3 -- \citet{Li2003}; 4 -- \citet{Abdo2009}; 5 -- \citet{Greiner2009};
6 -- \citet{Kumar2009}; 7 -- \citet{Ackermann2010}; 8 -- \citet{Kumar2010}.
\end{minipage}
\end{center}
\end{table*}

The data we use are listed in Table \ref{tab1}.  \citet[hereafter \citetalias{Panaitescu2002}]{Panaitescu2002}
have analyzed afterglow observations of ten GRBs, and
have derived for these objects the parameters we need. We make use of
the results in Tables 2 and 3 of their paper. We include GRB 970508
for which we take $E_\gamma$ from \citet{Bloom2003}, but
we omit GRB 980159 since the redshift is not known.  For the remaining
8 systems, we take \citetalias{Panaitescu2002}'s estimates of $E_\gamma$, the
beaming-corrected $\gamma$-ray emission in the 20--2000keV band, and
$E_0$, the beaming-corrected kinetic energy of the afterglow, and
compute the $\gamma$-ray efficiency parameter $\epsilon_\gamma$.  The
afterglow data considered by \citetalias{Panaitescu2002} did not include observations during
the early stages of the afterglow (first day or so).  During this
time, the external shock is expected to be somewhat radiative. It is
thus possible that \citetalias{Panaitescu2002} slightly underestimated $E_0$ in their models.
To allow for this, we double their values of $E_0$ and estimate the
$\gamma$-ray efficiency by $\epsilon_\gamma\sim
E_\gamma/(E_\gamma+2E_0)$.  (The correction factor of 2 is probably
too large, but our aim is to be conservative.)

In addition, we also estimate the magnetization $\sigma$ of the jet
ejecta.  Based on numerical and analytical work on cold
magnetically-accelerated GRB jets, \citet{Tchekhovskoy2010b} have
shown that the following inequality must be satisfied,
\begin{equation}
\gamma_j\sin\theta_j \lesssim 15\sigma^{1/2},
\end{equation}
where $\gamma_j$ and $\theta_j$ are the Lorentz factor and opening
angle of the jet ejecta at the conclusion of the prompt emission phase,
i.e., just before the onset of the afterglow phase.  The factor of
15 is a logarithmic term.  Since \citetalias{Panaitescu2002} have estimated $\gamma_j$ and
$\theta_j$ for the 9 GRBs of interest to us (see Table \ref{tab1}), from their
data we obtain for each GRB a lower limit on $\sigma$,
\begin{equation}
\sigma_{\rm min} =\left(\frac{\gamma_j\sin\theta_j}{15}\right)^2.
\label{sigmamin}
\end{equation}
The solid circles in Fig. \ref{fig1} show the values of $\sigma_{\rm min}$ and
$\epsilon_\gamma$ for the nine GRBs from the \citetalias{Panaitescu2002} sample.

In addition, we have gone through the literature and estimated
$\epsilon_\gamma$ and $\sigma_{\rm min}$ for three more bursts: GRB
021004 (relevant data taken from \citealt{Bloom2003};
\citealt{Li2003}), GRB 080916C \citep{Abdo2009,Greiner2009,Kumar2009}, 
GRB 090510 \citep{Ackermann2010,Kumar2010}. These three GRBs are shown in Fig. \ref{fig1} with open
circles, and the corresponding data are given in the last three lines
of Table \ref{tab1}.

In order to compare the data with the predictions of our internal
shock model, we need to assume a value for the electron heating
parameter $\epsilon_e$.  From modeling afterglow observations it is
possible to estimate $\epsilon_e$ for the forward shock in individual
GRBs (e.g., \citealt{Panaitescu2002}). The median value from a sample
of 39 GRB afterglows is $\epsilon_e\approx0.2$ (Santana \& Barniol Duran, in preparation).  The theoretical curves in the left panel of Fig. \ref{fig1}
correspond to this value of $\epsilon_e$.  A quick look shows that the
model completely misses the observations for all twelve GRBs in our
sample.  It has been known for some time that internal shocks
involving unmagnetized shocks cannot easily achieve the $\gamma$-ray
efficiencies required by observations \citep*{Kumar1999,Lazzati1999,Panaitescu1999}.  Magnetization
makes the problem worse. For the $\sigma$ values estimated for GRBs,
the predicted efficiency is much lower than for the unmagnetized case,
so the discrepancy is quite large. Note further that the values of
$\sigma$ plotted for the individual GRBs correspond to $\sigma_{\rm
min}$ (eq. \ref{sigmamin}). This means that the points might actually
lie even farther to the right, which would make the discrepancy
impossibly large.

The panel on the right in Fig. \ref{fig1} shows the highly optimistic case when
$\epsilon_e=1$. This might correspond, for example, to an
electron-positron jet. Even in this case, the majority of GRBs are
inconsistent with the magnetized internal shock model.  We thus
conclude that, if jet ejections are magnetized and are described by
ideal MHD, and if the blob Lorentz factors $\gamma$ and $\gamma_{\rm
ib}$ are not very different from the values considered in Fig. \ref{fig1} (note
that the largest value considered is $\gamma_{\rm ib}\sim200$ which is
very unlikely to be exceeded), then the magnetized internal shock
model is ruled out conclusively by the data.

\section{Reverse Shock Model}
\label{sec:reverseshockmodel}
\subsection{Solving the Jump Conditions}

In the case of the reverse shock, we have to consider four regions, as
described in \citet[hereafter \citetalias{Sari1995}]{Sari1995}:

\medskip\noindent
1. The external unshocked ISM, which is at rest in the lab frame
\newline\noindent
2. The shocked ISM
\newline\noindent
3. The shocked jet ejecta
\newline\noindent
4. The unshocked jet ejecta

\medskip\noindent Following \citetalias{Sari1995}, we use subscripts 1, 2, 3, 4 to
identify quantities in the four regions.  Regions 2 and 3 are in
pressure equilibrium across the contact discontinuity and move with
the same Lorentz factor. As measured in the lab frame, region 1 is at
rest, regions 2 and 3 move with Lorentz factor $\gamma_c$ ($c$ for
contact), and region 4 moves with the jet Lorentz factor $\gamma_j$.

In the analysis below we consider several distinct frames. First, we
have the lab frame in which the unshocked jet moves with Lorentz factor
$\gamma_j$ and the shocked gas moves with $\gamma_c$. Next, we have
the frame of the forward shock in which the unshocked and shocked ISM
move with Lorentz factors $\gamma_1$ and $\gamma_2$. Then, we have the
frame of the reverse shock in which the unshocked and shocked jet
ejecta move with Lorentz factors $\gamma_4$ and $\gamma_3$. Finally,
we have the frame of the shocked gas. This frame moves with respect to
the lab frame with a Lorentz factor $\gamma_c$, with respect to the
forward shock frame with a relative Lorentz factor $\gamma_2$, and
with respect to the reverse shock frame with a relative Lorentz factor
$\gamma_3$.

We consider regions 1 and 2 to be essentially unmagnetized and treat
the forward shock between these two regions as a hydrodynamic shock.
Furthermore, we assume that the relative Lorentz factor across this
shock is large, i.e., $\gamma_c, ~u_c\gg1$.  Let us transform into the
frame of the forward shock. The upstream gas is cold ($P_1=0$, recall
that subscript 1 refers to region 1), is unmagnetized (by assumption,
$\sigma=0$) and has a large Lorentz factor ($\gamma_1\gg1$). We can
use the results given in \S\ref{sec:A3} (with subscripts $u$, $d$ replaced by
$1$ and $2$) in the limit $\sigma\to 0$ to calculate the properties of
the shocked gas.  We then obtain the following standard results for
the downstream gas,
\begin{equation}
\beta_2=\frac{1}{3},\quad \gamma_2=\frac{3}{2\sqrt{2}},\quad
n_2=2\sqrt{2}\,n_1\gamma_1, \quad P_2=\frac{2}{3}\gamma_1^2n_1mc^2.
\end{equation}
The relative Lorentz factor between the two regions, which we call
$\gamma_c$, is equal to $\gamma_1/\sqrt{2}$. Thus, we find
\begin{equation}
n_2=4\gamma_cn_1, \qquad P_2=\frac{4}{3}\gamma_c^2n_1mc^2.\label{n2P2}
\end{equation}

Consider now the reverse shock between the magnetized regions 4 and 3.
The jet ejecta have a Lorentz factor $\gamma_j$ in the lab frame and a
magnetization $\sigma$. In the frame of the reverse shock, we do not
know a priory the value of the upstream Lorentz factor $\gamma_4$.
Therefore, as in \S\ref{sec:internalshockmodel}, we will solve for $\gamma_4$ via the jump
conditions (all the relations given in \S\ref{sec:relativisticperpendicularshock} are valid, except that
subscripts $u$ and $d$ should be replaced by 4 and 3, respectively),
plus an additional requirement. In \S\ref{sec:internalshockmodel}, the additional constraint was
the value of $\gamma$ (or equivalently $\gamma_{\rm ib}$).  Here, for
easy comparison with previous work in the literature, we will fit a
target value of the ``relativity'' parameter $\xi$ defined in \citetalias{Sari1995} and
\citet{Giannios2008b}. This parameter is less than unity for a
relativistic shock and greater than unity for a Newtonian shock.  In
Appendix \ref{appendixb} we show that for a magnetized flow
\begin{equation}
\xi = \left(\frac{R_{\rm dec}}{R_s}\right)^{1/2} = \left[\frac{3}{\gamma_j^2}
\,\frac{n_4}{n_1}\,(1+\sigma)\right]^{1/2},\label{xi}
\end{equation}
where $R_{\rm dec}$ and $R_s$ are the deceleration radius and the
spreading radius of the expanding ejecta.

Given the jet Lorentz factor $\gamma_j$ in the lab frame, the
magnetization of the jet material $\sigma$, and a target value of the
relativity parameter $\xi$, the calculation proceeds as follows. We
begin by guessing a value for $\gamma_4$, the Lorentz factor of the
upstream jet ejecta as viewed in the frame of the reverse shock.
Then, as described in \S\ref{sec:relativisticperpendicularshock}, we solve for all quantities in the
downstream region 3.  In the rest frame of the shocked gas, the gas
pressure is equal to
\begin{equation}
P_{\rm 3,gas}=n_3\theta_3mc^2 = n_4\frac{u_4}{u_3}\theta_3mc^2,
\end{equation}
and the magnetic pressure is equal to
\begin{equation}
P_{\rm 3,mag}=\frac{B_{\rm 3,rest}^2}{8\pi}=\frac{u_4^2}{\gamma_4^2u_3^2}
\,\frac{B_4^2}{8\pi} = n_4\frac{u_4^2}{2u_3^2}\sigma mc^2.
\end{equation}
Since regions 2 and 3 are in pressure balance, we thus obtain the
following condition,
\begin{equation}
P_{\rm 3,tot}=n_4\left(\frac{u_4}{u_3}\theta_3+
\frac{u_4^2}{2u_3^2}\sigma\right)mc^2 = P_2 =
\frac{4}{3}\gamma_c^2n_1mc^2.\label{n4n1}
\end{equation}
The Lorentz factor $\gamma_c$ on the right-hand side of equation
(\ref{n4n1}) is straightforward to calculate. In the reverse shock
frame, we know that region 4 has a Lorentz factor $\gamma_4$, whereas
its Lorentz factor in the lab frame is $\gamma_j$ (which is
given). Thus, we can calculate the Lorentz factor of the reverse shock
as seen in the lab frame (relativistic velocity subtraction). Once we
have this quantity, we can transform $\gamma_3$ (which is in the
reverse shock frame) to the lab frame to obtain $\gamma_c$, the
Lorentz factor of region 3 (as well as region 2) in the lab frame.

Having calculated $\gamma_c$, we obtain the density ratio $n_4/n_1$
from equation (\ref{n4n1}), and hence the value of $\xi$ from equation
(\ref{xi}). We check this against the target value of $\xi$, and
numerically adjust $\gamma_4$ until we achieve the value of $\xi$ we
seek.  At this point, we have the solution to the problem.

Once we have the solution, we can calculate the parameter
$\epsilon_\gamma$. As before, we will assume that a fraction
$\epsilon_e$ of the gas thermal enthalpy in region 3 is radiated.  As
measured in the lab frame, this corresponds to an energy per particle
of $\epsilon_e\gamma_c\, h(\theta_3)\theta_3 mc^2$.  To calculate the
total energy per particle of the system, it is simplest to consider
the unshocked jet fluid. In its own rest frame, the enthalpy per
particle is $mc^2(1+\sigma)$, and this gets multiplied by $\gamma_j$
when we transform to the lab frame.\footnote{Note that, for both of
the above quantities, we first calculate the enthalpy in the rest
frame of the gas, where there is no net momentum. Thus, transformation
of the energy to another frame requires only multiplication by the
relevant Lorentz factor.}  Thus we obtain
\begin{equation}
\epsilon_\gamma = \frac{\epsilon_e \gamma_c\, h(\theta_3)\theta_3}{\gamma_j\,
(1+\sigma)}.
\end{equation}

\begin{figure}
\includegraphics[width=\columnwidth]{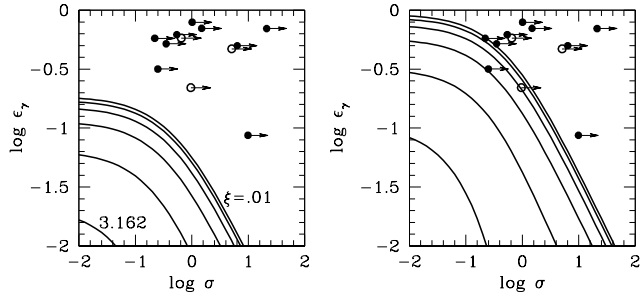}
\caption{Similar to Fig. \ref{fig1}, but for a magnetized reverse shock. The
curves correspond to jet Lorentz factor $\gamma_j=300$ as measured in
the lab frame, and (from below) relativity parameter $\xi=3.162$, 1,
0.3162, 0.1, 0.03162, 0.01. The panel on the left is for
$\epsilon_e=0.2$ and that on the right for $\epsilon_e=1$.}
\label{fig2}
\end{figure}

Apart from $\epsilon_e$, this reverse shock model has three
parameters: the jet Lorentz factor $\gamma_j$, the magnetization
parameter of the jet material $\sigma$, and the relativity parameter
$\xi$.  Figure \ref{fig2} shows results for a fixed value of $\gamma_j=300$
(the results hardly change for other values, e.g., 100 or 1000). The
curves correspond to selected values of the relativity parameter
$\xi$.  As in the case of unmagnetized reverse shocks (\citetalias{Sari1995},
\citealt{Giannios2008b}), we find that the $\gamma$-ray efficiency is highest for
highly relativistic shocks ($\xi\ll1$), and the efficiency is very
poor for Newtonian shocks ($\xi>1$).  In addition, for a given value
of $\xi$, the efficiency decreases as the magnetization increases.
This result is in qualitative agreement with the work of
\citet{Zhang2005}, \citet{Lyutikov2005} and \citet{Giannios2008b}.

\subsection{Comparison with GRB Data}

As in the case of the internal shock model, a comparison of the
predictions of the magnetized reverse shock model with GRB data
(Fig. \ref{fig2}) indicates that the model has no hope of satisfying the
observations.  For $\epsilon_e=0.2$, which we consider a reasonable
value, not a single GRB agrees with the model even if we assume an
extremely relativistic shock with $\xi=0.01$.  For $\epsilon_e=1$,
which in our opinion is rather optimistic, a few systems do fall
inside the model curves, but far too many systems still remain
unexplained.  We thus conclude that, if jet ejections are cold and
magnetized and are described by ideal MHD, then the magnetized reverse
shock model considered here is ruled out by the data.

\section{Summary and Discussion}
\label{sec:summaryanddiscussion}
The magnetic acceleration paradigm for relativistic jets is
theoretically appealing and widely accepted
(\citealt{Blandford1976,Lovelace1976,Begelman1994,Bogovalov1997,
Lyubarsky2001,Vlahakis2003}; see \citealt{Beskin2009} for a complete reference
list).  Recent advances in numerical techniques, coupled with
analytical methods, have led to a deeper understanding of how
Poynting-dominated jets accelerate to large Lorentz factors
\citep{Komissarov2004,Komissarov2007,Komissarov2009,Komissarov2010,Tchekhovskoy2008,Tchekhovskoy2009,Tchekhovskoy2010a,Tchekhovskoy2010b}.  
In the specific context of ultra-relativistic GRB jets, it has
recently been shown that the collimation angle $\theta_j$ and the
Lorentz factor $\gamma_j$ of a magnetically accelerated jet are not
independent but are related via the magnetization parameter $\sigma$
\citep{Tchekhovskoy2010b}.  Estimates of $\theta_j$ and $\gamma_j$ of
GRB jets, obtained by modeling afterglow data, are generally
consistent with these jets having $\sigma\sim1$ just prior to the
onset of the afterglow \citep{Tchekhovskoy2010b}.  This indicates
that GRB jets successfully convert about half of their initial
Poynting flux to matter kinetic energy by the time they reach the
deceleration radius.  These jets are thus energetically efficient.

When it comes to radiative efficiency, however, $\sigma\sim1$ is not
sufficient.  \citetalias{Kennel1984} showed that perpendicular shocks in cold magnetized
gas produce thermal energy very inefficiently unless $\sigma$ is much
less than unity.  We have explored this issue in detail in the context
of GRB internal shocks and reverse shocks. Both of these shocks occur
in the material ejected in a GRB and are expected to be magnetized.
The geometry is also such that the shocks will be perpendicular, i.e.,
the magnetic field will be perpendicular to the velocity vector, or
parallel to the shock front.  We have analyzed such shocks, assuming
that a fraction $\epsilon_e$ of the gas thermal energy in the shocked
gas goes into electrons and that this energy is entirely radiated in
prompt radiation.  We consider two values of $\epsilon_e$, viz.,
$\epsilon_e=0.2$, which we consider to be a reasonable estimate, and
$\epsilon_e=1$, which is highly optimistic.

Our calculations indicate that, once $\sigma$ exceeds about 0.1, the
efficiency of thermalization begins to fall noticeably, and that the
drop becomes quite precipitous once $\sigma>1$ (Figs. \ref{fig1}, \ref{fig2}).  GRB
observations indicate that the prompt $\gamma$-ray emission is quite
efficient, with the efficiency parameter $\epsilon_\gamma$ (defined in
eq. \ref{egamma}) being typically of order 0.5 or larger (Table \ref{tab1}). On the other
hand, not a single GRB has $\sigma<0.1$, as needed to obtain such high
efficiency in a cold magnetized shock, and half our sample has
$\sigma>1$, where radiative efficiency is very poor.  The implication
is that GRB prompt emission cannot be produced by either internal
shocks or the reverse shock, if jets are cold and magnetically
accelerated.  This conclusion is hard to avoid.  Even with very
optimistic assumptions, e.g., all the thermal energy goes into
electrons ($\epsilon_e=1$, which might happen if the jet is made
entirely of electrons and positrons), and is immediately radiated in
$\gamma$-rays, the calculated efficiency is far below what is needed
to explain the observations.

We consider here several possible resolutions of this puzzle, none of
which is very satisfactory.\footnote{\citet{Zhang2011} have
considered internal collisions for a magnetically dominated outflow
and suggest that this could facilitate dissipation of magnetic energy
via reconnection. We do not discuss this particular process here.}

\begin{figure}
\includegraphics[width=\columnwidth]{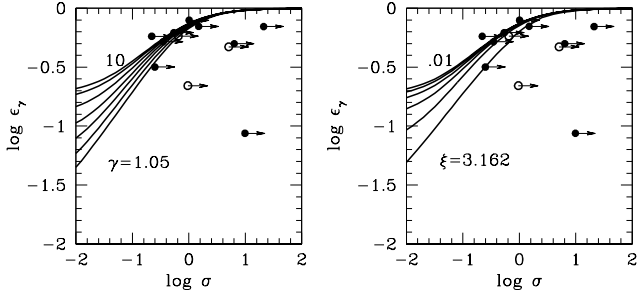}
\caption{Similar to Figs. \ref{fig1} and \ref{fig2}, except it is assumed that, in
addition to a fraction $\epsilon_e=0.2$ of the gas thermal enthalpy,
the entire magnetic enthalpy is also radiated as prompt $\gamma$-rays.
The panel on the left is for an internal shock and that on the right
for a reverse shock.}
\label{fig3}
\end{figure}

One possibility is to associate the prompt emission with the forward
shock, which is very weakly magnetized ($\sigma\ll1$) and therefore
converts a large fraction of the jet kinetic energy into thermal
energy.  This alone is not enough since the thermal energy must then
be radiated with greater than 50\% efficiency in order to explain the
observed values of $\epsilon_\gamma$.  Therefore, nearly all the
thermal energy should go into electrons ($\epsilon_e\sim1$), which is
very unlikely for the electron-proton plasma we expect to be present
in the forward shock. In any case, the forward shock has been
convincingly associated with afterglow emission (\citealt{Gehrels2009,
Meszaros2002,Piran2005,Zhang2007}; and references therein), and
it does not seem likely that the same region will also produce the
prompt emission.

A second possibility is that much of the magnetic energy in a GRB
shock is somehow converted to particle thermal energy. That is, when
$\sigma$ is large and most of the energy density in the post-shock gas
is in the form of magnetic energy, there is a mechanism whereby this
energy is converted to particle energy.  A scenario where this can
happen is if the pre-shock gas is ``striped'' as in current models of
energy dissipation in the magnetized wind of pulsars
\citep{Lyubarsky2001,Lyubarsky2010a}.  
A striped morphology is not obvious for
a GRB jet (but see \citealt{McKinney2010}). However, if it is
present, we do expect a substantial fraction of the magnetic energy to
be dissipated in the shock. Figure \ref{fig3} shows results for a hypothetical
model in which we assume that, in addition to a fraction
$\epsilon_e=0.2$ of the gas thermal enthalpy, 100\% of the magnetic
energy is radiated.  This model does explain the GRB observations but
at the price of making a very extreme (and theoretically unsupported)
assumption. We do not endorse this model but present it merely as a
way to emphasize how difficult it is to explain the radiative
efficiency of GRB prompt emission.

A third possibility is that the magnetic energy is dissipated, not
through a shock, but through some other ``current-driven'' mechanism
such as reconnection.  Poynting-dominated magnetically accelerated
jets are fairly stable once they are ultra-relativistic (e.g.,
\citealt{NarayanLiTchekhovskoy2009}) and are unlikely to have
large-amplitude fluctuations that might drive reconnection. However,
it is conceivable that these jets lose their stability once they reach
a large radius \citep{McKinney2010}, e.g., the deceleration
radius where the jet meets the external medium and begins to slow
down.  Whether the instability would be powerful enough to drive
wholesale reconnection and convert most of the magnetic energy into
particle energy is an open question. As Fig. \ref{fig3} shows, something like
this is needed if one is to explain the data.

Another possibility is that our assumption of cold gas, whose
acceleration is entirely by magnetic means, is incorrect.
Non-relativistic MHD simulations of magnetized jets \citep*{Moll2008,Moll2009} indicate that these jets develop a kink
instability which might lead to dissipation.  We could then have a
scenario in which the jet starts off magnetically dominated at the
base but quickly dissipates its magnetic energy into heat while the
jet is still non- or quasi-relativistic.  Further acceleration of the
jet is then driven by the thermal pressure of the heated gas.  Thus,
we no longer have a magnetically driven jet, but something akin to the
standard fireball model of a GRB.  Clearly, the calculations presented
here, which are restricted to cold magnetized gas, are not relevant
for such a model.

Finally, it is possible that the prompt emission in GRBs is not
produced in the jet at a large distance from the progenitor, but
rather in the photospheric region where the jet ejecta first become
transparent.  Models of this form have been developed \citep{Thompson1994,
Meszaros2000, Rees2005,Peer2006,Giannios2008a,Beloborodov2010,Metzger2011} and it is claimed that they produce prompt
$\gamma$-ray emission with high radiative efficiency and with the
correct spectrum (e.g., \citealt*{Peer2011,Vurm2011}).  Magnetic fields may play a role in photospheric
models \citep{Uzdensky2010}, but the role of shocks is
unclear.  Our analysis is not applicable to these models.

\section*{Acknowledgments}
RN and AT were supported in part by NASA grant NNX11AE16G and NSF
grant AST-1041590, and by NSF through TeraGrid resources provided by
QueenBee of the Louisiana Optical Network Initiative 
(\href{http://www.loni.org}{http://www.loni.org}) and Kraken of the National 
Institute for Computational Sciences 
(\href{http://www.nics.tennessee.edu}{http://www.nics.tennessee.edu})
under grant number TG-AST080026N (RN \& AT) and TG-AST100040 (AT). AT was
supported in part by a Princeton Center for Theoretical Science Fellowship. PK was
supported in part by NSF grant AST-0909110.

\appendix

\section{Analytical Approximations for a Perpendicular Shock}
\label{appendixa}

A cold hydrodynamic flow can have a shock for any choice of the
upstream velocity $u_u$. This is because no signals can propagate in
the cold gas and so the upstream gas is always supersonic. A magnetized
fluid is different. Even if the gas is cold, Alfven and fast
magnetosonic waves can still propagate in the fluid.  Thus a shock is
possible only if the upstream gas moves faster than these waves.

For the particular geometry we have considered, viz., a perpendicular
shock with magnetic field perpendicular to the velocity vector, the
relevant wave speed is that of the fast magnetosonic wave, which is
given (in the comoving frame of the gas) by
\begin{equation}
\beta_{\rm fms}^2=\frac{\sigma}{\sigma+1}, \quad
u_{\rm fms}^2=\sigma.
\end{equation}
Thus we can have a shock only if
\begin{equation}
u_u^2 > \sigma.
\end{equation}

We now consider a number of limiting cases. When $u_u^2 \gg \sigma$,
we expect to have a strong shock, whereas when $u_u^2$ is only
marginally greater than $\sigma$, we expect a weak shock.  In
addition, we have different results depending on whether $u_u^2 \gg 1$
(ultrarelativistic) or $u_u^2\ll 1$ (nonrelativistic), and on whether
$\sigma \gg 1$ (strongly magnetized) or $\sigma \ll 1$ (weakly
magnetized).  \citetalias{Kennel1984} considered a couple of important cases, but here we
present scalings for all the different regimes. Figure \ref{fig4} identifies
the regimes and labels them by the respective subsection where each
is discussed.

\begin{figure}
\centering
\includegraphics[width=\columnwidth]{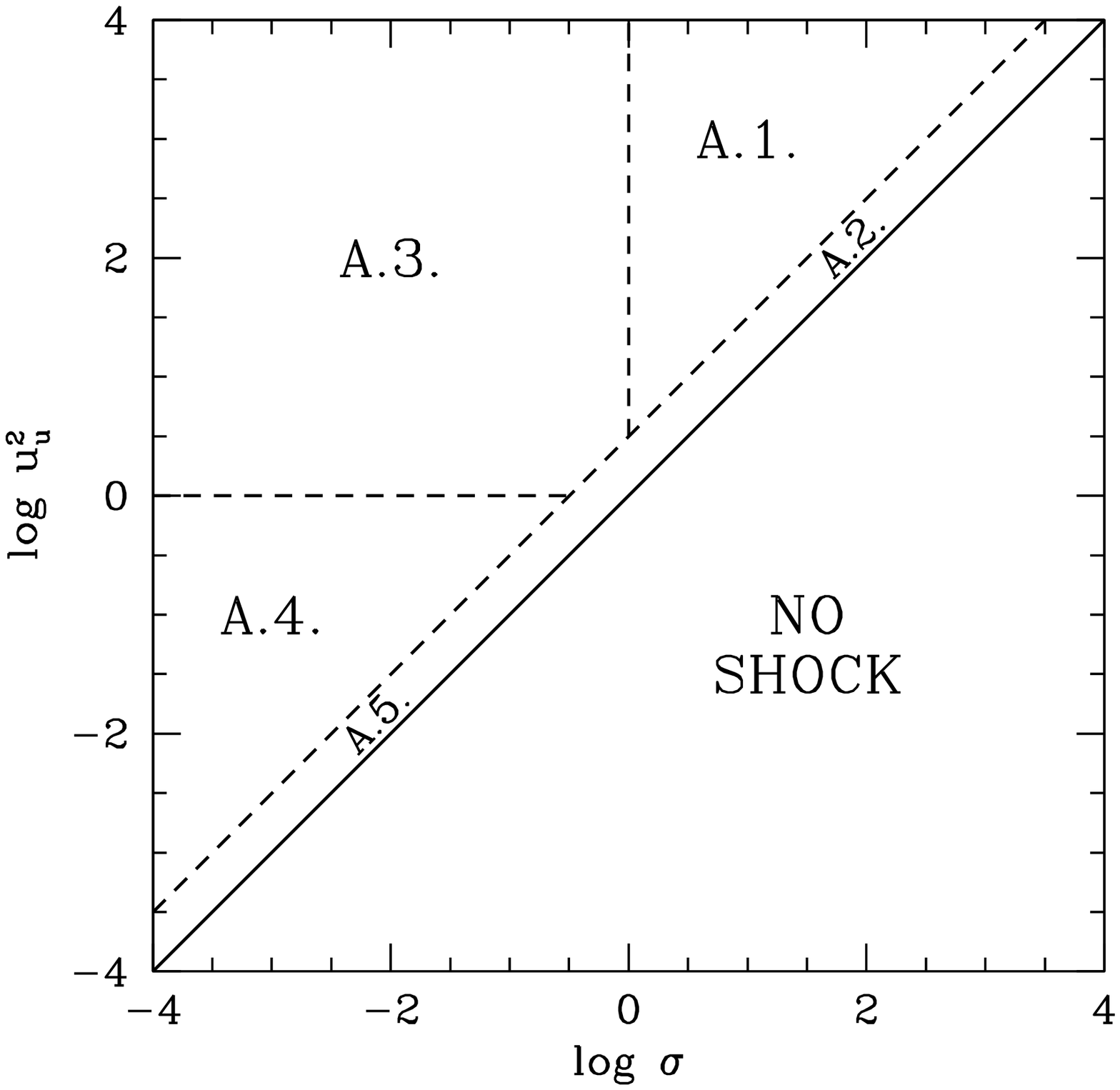}
\caption{Shows different regimes for a perpendicular shock in cold
upstream gas.  The horizontal axis indicates the magnetization
parameter $\sigma$ of the upstream gas (eq. \ref{sigma}) and the vertical axis
indicates the square of the upstream relativistic velocity $u_u$ as
measured in the frame of the shock.  There is no shock when $u_u
<\sqrt{\sigma}$ since the upstream gas moves slower than the fast
magnetosonic wave speed $u_{\rm fms}=\sqrt{\sigma}$ (eqs. A1, A2).
The regions marked A.1.,...,A.5, in the plot correspond to different
shock regimes. The labels refer to the section numbers in Appendix \ref{appendixa}
where each is discussed. Regions A.1. and A.2. correspond to highly
magnetized shocks ($\sigma\gg1$), which are necessarily also highly
relativistic ($u_u\gg1$).  A.1. describes a strong shock, where $u_u
\gg \sqrt{\sigma}$, while A.2. describes a weak shock where $u_u$ is
only marginally greater than $\sqrt{\sigma}$.  Regions A.3., A.4.,
A.5., correspond to weakly magnetized shocks ($\sigma\ll1$).  A.3.  is
highly relativistic ($u_u\gg1$), where the shock is necessarily
strong. A.4. is non-relativistic ($u_u\ll1$), but still with a strong
shock, while A.5. corresponds to a non-relativistic weak shock, where
$u_u$ is only marginally greater than $\sqrt{\sigma}$.}
\label{fig4}
\end{figure}

\subsection{Strong Relativistic Shock with Strong Magnetization: 
$u_u^2 \gg \sigma \gg 1$}
\label{sec:A1}
This case has been considered by \citetalias{Kennel1984} who give the following results:
\begin{align}
u_d^2 &\approx \sigma +\frac{1}{8}+\frac{1}{64\sigma}+\os{\frac{1}{\sigma}}, \\
\frac{B_d}{B_u} &\approx 1+\frac{1}{2\sigma}-\frac{3}{16\sigma^2}+\os{\frac{1}{\sigma^2}}, \\
\frac{\theta_d}{u_u} &\approx \frac{1}{8\sqrt{\sigma}}
\left(1-\frac{3}{16\sigma}\right)+\os{\frac{1}{\sigma^{3/2}}},
\end{align}
where $\os{x}$ denotes terms of higher order than $x$, i.e.,
$\os{x}/x\to0$ as $x\to0$.
In the above, we remind the reader that the magnetic field strengths $B_u$ and
$B_d$ are measured in the shock frame. These results are obtained by
expanding the jump conditions as power series in the small quantity
$1/\sigma$, and matching terms of similar order.  We have set
$h(\theta_d)=4$, as appropriate for relativistically hot downstream
gas.  Our result for $\theta_d/u_u$ differs from that given in \citetalias{Kennel1984}.

\subsection{Weak Relativistic Shock with Strong Magnetization: 
$u_u^2\to\sigma \gg 1$}
\label{sec:A2}

As $u_u^2$ approaches $\sigma$, the shock becomes progressively
weaker. Let us write $u_u^2=\sigma (1+\Delta)$, with $\Delta \ll 1$.
As $\Delta$ becomes progressively smaller, less and less of the
kinetic energy of the upstream gas is thermalized in the shock. In
this limit, the downstream temperature $\theta_d$ becomes
non-relativistic and so we set $h(\theta_d)=5/2$. In this limit, the
leading terms in the solution are as follows:
\begin{align}
u_d^2 &\approx \sigma\left[1-\frac{\Delta}{3}+\frac{44}{81}
\left(1-\frac{3}{11\sigma}\right)\Delta^2\right]+\os{\Delta^2}, \\
\frac{B_d}{B_u} &\approx 1+\frac{2}{3\sigma}\left(1-\frac{1}{\sigma}\right)\Delta
-\frac{58}{81\sigma}\Delta^2+\os{\frac{\Delta^2}{\sigma};\frac{\Delta}{\sigma^2}}, \\
\theta_d &\approx \frac{4}{81}\left(1-\frac{1}{\sigma}+
\frac{1}{\sigma^2}\right)\Delta^3-\frac{8}{81}\left(1-
\frac{4}{3\sigma}\right)\Delta^4+\os{\frac{\Delta^3}{\sigma^2};\frac{\Delta^4}{\sigma}}.
\end{align}
Note that the leading term in the temperature of the shocked gas goes
as $\Delta^3$, i.e., the shock is extremely inefficient. This is a
characteristic feature of weak shocks.

\subsection{Strong Relativistic Shock with Weak Magnetization: 
$u_u^2 \gg 1 \gg \sigma$}
\label{sec:A3}
This case has again been considered by \citetalias{Kennel1984}. With $h(\theta_d)=4$, the
results are
\begin{align}
u_d^2 &\approx \frac{1}{8}+\frac{9}{8}\sigma -\frac{9}{8}\sigma^2+\os{\sigma^2}, \\
\frac{B_d}{B_u} &\approx 3-12\sigma+96\sigma^2+\os{\sigma^2}, \\
\frac{\theta_d}{u_u} &\approx \frac{1}{3\sqrt{2}}
\left(1-\frac{5}{2}\sigma+\frac{111}{8}\sigma^2\right)+\os{\sigma^2}.
\end{align}
Again, our result for $\theta_d$ differs from that given in \citetalias{Kennel1984} 
The results for an unmagnetized relativistic shock are
recovered by simply setting $\sigma=0$ in the above relations.

\subsection{Strong Non-Relativistic Shock with Weak Magnetization: 
$1 \gg \beta_u^2 \gg \sigma$}
\label{sec:A4}

Here we consider the non-relativistic case and replace $u_u, ~u_d$ by
$\beta_u, ~\beta_d$. Also, we set $h(\theta_d)=5/2$. Then we find
\begin{align}
\beta_d &\approx \frac{1}{4}\beta_u+\frac{9}{8}\frac{\sigma}{\beta_u}
-\frac{9}{4}\frac{\sigma^2}{\beta_u^3}+\os{\frac{\sigma^2}{\beta_u^3}}, \\
\frac{B_d}{B_u} &\approx 4-18 \frac{\sigma}{\beta_u^2}
+117\frac{\sigma^2}{\beta_u^4}+\os{\frac{\sigma^2}{\beta_u^4}}, \\
\theta_d &\approx \frac{3}{16}\beta_u^2-\frac{21}{16}\sigma+\os{\sigma}.
\end{align}
The solution for an unmagnetized shock is obtained by setting
$\sigma=0$.  As an aside, we note that $\sigma$ is related to the
Alfven wave speed $v_A$ by
\begin{equation}
\sigma = \frac{B^2}{4\pi nmc^2} = \frac{v_A^2}{c^2} = \beta_A^2.
\end{equation}

\subsection{Weak Non-Relativistic Shock with Weak Magnetization: 
$1 \gg \beta_u^2 \to \sigma$}
\label{sec:A5}

Finally, we consider the case when the shock is non-relativistic and
$\beta_u = \sqrt{\sigma}(1+\Delta)$ with $\sqrt\sigma\ll\Delta \ll 1$. In this
limit, we find
\begin{align}
\beta_d &\approx \left(1-\frac{1}{3}\Delta+\frac{46}{81}
\Delta^2\right)\sqrt{\sigma}+\os{\Delta^2\sqrt\sigma}, \\
\frac{B_d}{B_u} &\approx 1+\frac{4}{3}\Delta-\frac{10}{81}\Delta^2+\os{\Delta^2}, \\
\theta_d &\approx \left(\frac{32}{81}\Delta^3-\frac{112}{243}\Delta^4
\right)\sigma+\os{\Delta^4\sigma}.
\end{align}

\section{The Relativity Parameter $\xi$}
\label{appendixb}

We generalize the discussion of the parameter $\xi$ given in \citetalias{Sari1995},
following the analysis of \citet{Giannios2008b}.  We consider a
spherically expanding shell of cold magnetized jet material of radius
$R$, shell thickness $\Delta$, and Lorentz factor $\gamma_j$, all
measured in the lab frame.  The ``spreading radius'' of the shell is
given by
\begin{equation}
R_s=\gamma_j^2\Delta.
\end{equation}

The total (isotropic equivalent) energy of the shell is
\begin{equation}
E=4\pi R^2\Delta n_4mc^2\gamma_j^2(1+\sigma) \equiv
M_{\rm ej}\gamma_j c^2(1+\sigma),
\end{equation}
where $n_4$ is the rest frame particle number density of the jet
material, $\sigma$ is the magnetization of the material, and $M_{\rm
ej}$ is the total rest mass of the shell. From $E$, we obtain the
``Sedov length'' $\ell$ and the ``deceleration radius'' $R_{\rm dec}$,
\begin{eqnarray}
\ell &=& \left(\frac{3E}{4\pi n_1mc^2}\right)^{1/3} =
\left[\frac{3M_{\rm ej}\gamma_j(1+\sigma)}{4\pi n_1m}\right]^{1/3}, \\
R_{\rm dec} &=& \frac{\ell}{\gamma_j^{2/3}}=
\left[\frac{3M_{\rm ej}(1+\sigma)}{4\pi n_1m\gamma_j}\right]^{1/3},
\end{eqnarray}
where $n_1$ is the number density of the external ambient medium.
Substituting for $M_{\rm ej}$ (with $R=R_{\rm dec}$) in the equation
for $R_{\rm dec}$, we find that
\begin{equation}
R_{\rm dec} = 3\Delta\,\frac{n_4}{n_1}\,(1+\sigma),
\end{equation}
from which we obtain
\begin{equation}
\xi = \left(\frac{R_{\rm dec}}{R_s}\right)^{1/2} =
\left[\frac{3}{\gamma_j^2}\,\frac{n_4}{n_1}\,(1+\sigma)\right]^{1/2}.
\end{equation}
Note that $n_4$ is the number density of the jet ejecta at the moment
when the shell radius $R$ is equal to $R_{\rm dec}$.

\label{lastpage}
\end{document}